\documentclass[11pt,twoside]{book}
\usepackage{asp2010}
\usepackage{natbib}
\usepackage{graphicx}
\usepackage{multirow}
\begin{document}
\title{A Roe-type Riemann solver based on the spectral decomposition
  of the equations of Relativistic Magneto\-hydrodynamics}

\author{
Jos\'e M$^{\underline{\mbox{a}}}$ Ib\'a\~nez\altaffilmark{1}, 
Miguel A. Aloy\altaffilmark{1}, Petar Mimica\altaffilmark{1}, Luis Ant\'on\altaffilmark{1},\\
 Juan A. Miralles\altaffilmark{2}, Jos\'e M$^{\underline{\mbox{a}}}$ Mart\'{\i}\altaffilmark{1}}

\altaffiltext{1}{Department of Astronomy and Astrophysics,
University of Valencia, 46100 Burjassot (Valencia), Spain}
\altaffiltext{2}{Department of Applied Physics,
University of Alicante, Ap. Correus 99,
03080 Alacant, Spain}

\begin{abstract}
%  In a recent paper (Ant\'on et al. 2010) we have derived sets of
  In a recent paper \citep{Antonetal10} we have derived sets of
  right and left eigenvectors of the Jacobians of the relativistic MHD
  equations, which are regular and span a complete basis in any
  physical state including degenerate ones. We present a summary of
  the main steps followed in the above derivation and the numerical
  experiments carried out with the linearized (Roe-type) Riemann
  solver we have developed, and some note on the (non-)convex
  character of the relativistic MHD equations.

  \end{abstract}

%        %%%%%%%%%%%%
\section{Introduction}
\label{intro}
%        %%%%%%%%%%%%

Relativistic flows in association with intense gravitational and
magnetic fields are commonly linked up to extremely energetic
phenomena in the Universe, viz. pulsar winds, anomalous X-ray pulsars,
soft gamma-ray repeaters, gamma-ray bursts, relativistic jets in
active galactic nuclei, etc. The necessity to model the aforementioned
astrophysical scenarios in the framework of relativistic MHD (RMHD),
together with the fast increase in computing power, is pushing towards
the development of more efficient numerical algorithms. In the
last years, considerable progress has been achieved in numerical
special RMHD (SRMHD), by extending the existing high-resolution
shock-capturing (HRSC) methods of special relativistic hydrodynamics
\citep[e.g.,][]{martilr:03}.  In the so called Godunov-type methods,
an important subsample of HRSC methods, numerical fluxes are evaluated
through the exact or approximate solution of the (local) Riemann
problem.  Despite the fact that such an exact solution in SRMHD
is known \citep{romero,GR06}, approximate algorithms are usually
preferred because of their larger numerical efficiency. Several
%authors \citep[see, e.g.,][and references therein]{Anton:2010} have
authors \citep[see, e.g.,][and references therein]{Antonetal10} have
developed independent {\it Roe-type} algorithms based on linearized
Riemann solvers relying on the characteristic structure of the RMHD
equations.

% Our goal

The purpose of the present paper is twofold. On one hand, the
objective is to present a {\it regular} set of right and left
eigenvectors of the flux vector Jacobian matrices of the RMHD equations, and
span a complete basis in {\it any} physical state, including
degenerate states. On the other hand, wish to evaluate numerically the
performance of a RMHD Riemann solver based on the aforementioned
spectral decomposition. Both the theoretical analysis and the
numerical applications presented in this paper are based on the work
%developed by \cite{Anton:2010}, where we have characterized thoroughly
developed by \cite{Antonetal10}, where we have characterized thoroughly
all the degeneracies of RMHD in terms of the components of the
magnetic field normal and tangential to the wavefront in the fluid
rest frame. Our numerical method deviates in several aspects from
previous works based on linearized Riemann solver approaches
\citep{komissarov99,Balsara01,koldoba}. First, numerical fluxes are
computed from the spectral decomposition in conserved
variables. Second, we present explicit expressions also for the left
eigenvectors. Third, and most important, we have extended classical
MHD strategy \citep{BW88} to relativistic flows, giving sets of right
and left eigenvectors which are well defined through degenerate
states. Based on the full wave decomposition (FWD) provided by the
renormalized set of eigenvectors in conserved variables, we have also
developed a linearized (Roe-type) Riemann solver.

Extensive testing against one- and two-dimensional standard numerical
problems allows us to conclude that our solver is very robust. When
compared with a family of simpler solvers that do not require the
knowledge of the full characteristic structure of the equations in the
computation of the numerical fluxes, our solver turns out to be less
diffusive than HLL and HLLC, and comparable in accuracy to the HLLD
solver. The amount of operations needed by the FWD solver makes it
less efficient computationally than those of the HLL family in
one-dimensional problems.  However its relative efficiency increases
in multidimensional simulations.

%        %%%%%%%%%%%%%%%%%%%%%%%%%%%%%%%%%%%%%%%%%%%%%%%%%%%%%%%%
\section{The equations of ideal relativistic magnetohydrodynamics}
%        %%%%%%%%%%%%%%%%%%%%%%%%%%%%%%%%%%%%%%%%%%%%%%%%%%%%%%%%

The equations of ideal RMHD correspond to the conservation of
rest-mass and energy-momentum, and the Maxwell equations. In the
following, the standard Einstein sum convention is assumed. Greek
indices will run from 0 to 3 (or from $t$ to $z$) while Roman run from
1 to 3 (or from $x$ to $z$). We use units in which the speed of light
is $c=1$ and the $(4 \pi)^{1/2}$ factor is absorbed in the definition
of the magnetic field. Specializing for a flat space-time and
Cartesian coordinates, these equations can be written as a system of
conservation laws, which reads
%where subscript $_{,\mu}$ denotes partial derivative with respect to the corresponding coordinate, $(t,x,y,z)$, and 

\begin{equation}
\frac{\partial {\bf U}}{\partial t} +
\frac{\partial {\bf F}^{i}}{\partial x^{i}} = 0,
\label{eq:fundsystem}
\end{equation}
\noindent
where the state vector, ${\bf U}$ (vector of {\it conserved variables}), 
and the fluxes, ${\bf F}^i$ ($i=1,2,3$ 
or $i=x,y,z$), are the following column vectors,
\begin{equation}
{\bf U}  =   (D,  S^j,   \tau,  B^k)^T
\label{state_vector}
\end{equation}
\begin{equation}
{\mathbf F}^i  = (D v^i, S^j v^i + p^{*} \delta^{ij} - b^j B^i/W, 
 \tau v^i + p^{*} v^i - b^0 B^i/W ,  
 v^i B^k - v^k B^i )^T
\label{flux2}
\end{equation}
\noindent
where the superscript $^{\rm T}$ stands for the transposition. 

  In the preceding equations, $D$, $S^j$ and $\tau$ stand, respectively, 
for the rest-mass density, the momentum density of the magnetized
fluid in the $j$-direction and its total energy density as measured in 
the laboratory (i.e., Eulerian) frame,
\begin{equation}
%\label{eq:D}
 D = \rho W
\,,\,\,\,\,\,\,\,\,\,
%\label{eq:Sj}
  S^j = \rho h^* W^2 v^j - b^0 b^j
\,,\,\,\,\,\,\,\,\,\,
%\label{eq:tau}
  \tau = \rho h^* W^2 - p^* - (b^0)^2 - D.
\end{equation}
\noindent
where $\rho$ is the proper rest-mass density, $h^* =1 + \epsilon +
p/\rho + b^2/\rho$ is the specific enthalpy including the contribution
from the magnetic field ($b^2$ stands for $b^\mu b_\mu$), $\epsilon$
is the specific internal energy, $p$ the thermal pressure, and $p^* =
p + b^2/2$ the total pressure.
The four-vectors representing the fluid velocity, $u^\mu$, and the
magnetic field measured in the fluid rest frame, $b^\mu$, and there is
an equation of state relating the thermodynamic variables, $p$, $\rho$
and $\epsilon$, $p = p(\rho, \epsilon)$. All the discussion will be
valid for a general equation of state but results will be shown for an
ideal gas, for which $p = (\gamma - 1) \rho \epsilon$, where $\gamma$
is the adiabatic exponent.  Quantities $v^i$ stand for the components
of the fluid velocity trivector as measured in the laboratory frame;
they are related with the components of the fluid four-velocity
according to the following expression $u^\mu = W(1, v^x, v^y, v^z)$,
where $W$ is the flow Lorentz factor, $W^2=1/(1-v^i v_i)$.

  The following fundamental relations hold between the components of the magnetic field four-vector in the comoving frame and the three vector components $B^i$ measured in the laboratory frame,

\begin{equation}
%\label{b0}
  b^0  =  W\, {\bf B} \cdot {\bf v} 
\,\,\,\,\,,\,\,\,\,\,
% \label{bi} 
  b^i  =  \frac{B^i}{W} + b^0 v^i 
\end{equation}
${\bf v}$ and ${\bf B}$ being, respectively, the tri-vectors $(v^x,v^y,v^z)$ and $(B^x,B^y,B^z)$.
\begin{equation}
  b^2 = \frac{{\bf B}^2}{W^2} + ({\bf B} \cdot {\bf v})^2 \ .
\end{equation}
  The preceding system must be complemented with the usual divergence constraint
\begin{equation}
  \label{eq:divb}
  \frac{\partial B^i}{\partial x^i} = 0\;, 
\end{equation}
\noindent
which should be fulfilled at all times.

  Fluxes ${\mathbf F}^i$ ($i = x,y,z$) are functions of the conserved variables, 
${\mathbf U}$, although for the RMHD this dependence, in general, can not be 
expressed explicitly. It is therefore necessary to introduce another set of variables, 
the so-called {\it primitive variables}, derived from the conserved ones, in terms of 
which the fluxes can be computed explicitly. We have used the following set of 
primitive variables 
\begin{equation}
\label{primitive-var}
  {\bf V} = (\rho, p, v^x, v^y, v^z, B^x, B^y, B^z)^T.
\end{equation}

%        %%%%%%%%%%%%%%%%%%%%%%%%%%%%%%%%%%%%%%%%%%%%%%
\section{Characteristic structure of the RMHD equations} \label{s:csrmhde}
%        %%%%%%%%%%%%%%%%%%%%%%%%%%%%%%%%%%%%%%%%%%%%%%

The hyperbolicity of the equations of RMHD including the derivation of
wavespeeds and the corresponding eigenvectors, and the analysis of
various degeneracies has been reviewed by \cite{anile}, in a covariant
framework, using a set of variables of dimension 10, the so-called
{\it covariant variables} (Anile's variables, in the next):

\begin{equation}
  {\bf \tilde{U}} = (u^\mu, b^\mu, p, s)^T,
\label{Anile-var}
\end{equation}

\noindent
where $s$ is the specific entropy.

  In terms of variables ${\bf \tilde{U}}$, the system 
of RMHD equations can be written as a quasi-linear system of the form

\begin{equation}
  {\cal A}^\mu {\bf \tilde{U}}_{; \mu}= 0, 
\label{amuab}
\end{equation}

\noindent
where the subscript ${; \mu}$ stands for the covariant derivative, and 
four $10\times 10$ Jacobian matrices ${\cal A}^{\mu}$ can be
found in Anile's book. 
It is important to remark that the 10 covariant variables we have 
used to write the system of equations are not independent, since 
they are related by the constraints 
\begin{equation}
 u^{\alpha} u_{\alpha} = -1
\,\,\,\,\,,\,\,\,\,\,
b^{\alpha} u_{\alpha} = 0
\,\,\,\,\,,\,\,\,\,\,
\partial_\alpha (u^\alpha b^0 - u^0 b^\alpha) = 0\,,
\end{equation}
\noindent
The latter condition, is a covariant representation of the divergence
constraint (Eq.~\ref{eq:divb}).

%           ------------------------------
\subsection{Wavespeeds and degeneracies}
%           ------------------------------
\label{ss:ws}

The system of (ideal) RMHD equations have the same seven wavespeeds as
in classical MHD: the entropic, Alfv\'en, slow magnetosonic, and fast
magnetosonic waves. They can be ordered as follows
\begin{equation}
\label{order}
\lambda^-_f \le \lambda^-_a \le \lambda^-_s \le \lambda_e \le \lambda^+_s \le \lambda^+_a \le \lambda^+_f,
\end{equation}
\noindent
where the subscripts $e$, $a$, $s$ and $f$ stand for {\it entropic}, 
{\it Alfv\'en}, {\it slow magnetosonic} and {\it fast magnetosonic} respectively, 
and the superscript $-$ or $+$ refer to the lower or higher value of each pair. 
Unlike classical MHD, it is however not possible, in general, 
to obtain simple expressions for the magnetosonic speeds since they are given 
by the solutions of a quartic equation.

%\subsection{Degeneracies}
%           ------------
%\label{ss:degs}

As in the case of classical MHD, degeneracies are encountered for
waves propagating perpendicular to the magnetic field direction (Type
I) and for waves propagating along the magnetic field direction (Type
II).  Finally, a particular subcase of Type II degeneracy appears when
the sound speed is equal to $c_a$.

  For Type I degeneracy, the two Alfv\'en waves, the entropic wave and 
the two slow magnetosonic waves propagate at the same speed 
($\lambda_a^-=\lambda_s^-= \lambda_e=\lambda_s^+=\lambda_a^+$). 
For Type II degeneracy, an Alfv\'en wave and a magnetosonic wave 
(slow or fast) of the same class propagate at the same speed 
($\lambda_f^-=\lambda_a^-$ or $\lambda_a^-=\lambda_s^-$ or 
$\lambda_s^+=\lambda_a^+$ or $\lambda_a^+=\lambda_f^+$). In the special 
Type II$^\prime$ subcase, an Alfv\'en wave and both the slow and fast 
magnetosonic waves of the same class propagate at the same speed 
($\lambda_f^-=\lambda_a^-=\lambda_s^-$ or $\lambda_s^+=\lambda_a^+=\lambda_f^+$). 

\subsection{Renormalized right eigenvectors}
%           -------------------------
\label{ss:renom_eigen}

As it is well known in classical MHD, Alfv\'en and magnetosonic
eigenvectors have a pathological behaviour at degeneracies, since they
become zero or linearly dependent and they do not form a basis.  In
%\cite{Anton:2010}, we have derived a new set of renormalized Alfv\'en
\cite{Antonetal10}, we have derived a new set of renormalized Alfv\'en
and magnetosonic eigenvectors for RMHD.  Our renormalized Alfv\'en
right eigenvectors (following \citealt{BW88} methodology) are a linear
combination of the ones proposed by \cite{komissarov99}, for the Type
II degeneracy case. However, contrary to the Komissarov's choice, our
expressions are free of pathologies not only in the Type II degeneracy
but also in the Type I degeneracy case.  Our derivation of the
renormalized magnetosonic right eigenvectors is algebraically more
cumbersome and reader interested is addressed to
%\cite{Anton:2010}. The final result of this analysis allows to have a
\cite{Antonetal10}. The final result of this analysis allows to have a
complete set of right eigenvectors linearly independent for all
possible states. Following the same procedure we have used to
renormalize the right eigenvectors, we have derived
%\cite[see][]{Anton:2010} left eigenvectors well behaved for degenerate
\cite[see][]{Antonetal10} left eigenvectors well behaved for degenerate
states.
%
%---------------------------------------------------------------------------
\section{A Full Wave Decomposition Riemann Solver in RMHD (FWD)}
%---------------------------------------------------------------------------
%
Let us summarize the main steps allowing us to derive a FWD Riemann
Solver in RMHD:
\begin{itemize}
\item Let $r_{_{\widetilde{\bf U}}} $ be a generic right
eigenvector derived in terms of Anile's variables (\ref{Anile-var}).
\item The corresponding eigenvector in terms of the primitive variables
(\ref{primitive-var}) is derived according to: $ \displaystyle{
  r_{_{\bf V}}= \left( \partial_{\widetilde{\bf U}} {\bf V}
  \right) r_{_{\widetilde{\bf U}}} }$. 
\item Finally, the corresponding vector in terms of the
  conserved variables (\ref{state_vector}) is obtained from $
  \displaystyle{ {\bf R}\equiv r_{_{\bf U }}= \left( \partial_{\bf V}
      {\bf U } \right) r_{_{\bf V}} } $.  
\item Analogously, for the left eigenvectors. This procedure,
  which starts with renormalized eigenvectors, allows one to get the
  full spectral decomposition of the Jacobian matrices (associated to
  the fluxes), and free of pathologies in the degeneracies.
\item We use a linearized (Roe's type) Riemann solver:\\
$
          \widehat{{\bf f}}_{j\pm{1\over 2}} = \frac{1}{2}
          \left( {\bf f}({\bf u}_{j\pm{1\over 2}}^{L})  +
          {\bf f}({\bf u}_{j\pm{1\over 2}}^{R}) -
          \sum_{\alpha = 1}^{p} \mid \widetilde{\lambda}_{\alpha}\mid
          \Delta \widetilde {\omega}_{\alpha}
          \widetilde {r}_{\alpha}
\right)$
\\
\noindent
where ${\bf u}^L$, ${\bf u}^R$, are the left and right reconstructed
variables; $\Delta {\omega}$, is the jump of characteristic variables.
\end{itemize}

Our FWD linearized Riemann solver has been exhaustively tested
%\citep{Anton:2010}.
\citep{Antonetal10}.

\section{A note on RMHD convexity}
%            %%%%%%%%%%%%%%%%%%%%%%%%%%%%%%%%%%%%
\label{s:convex}

For the sake of conciseness let us remind some definitions. A
characteristic field ${\cal C}_{\alpha} \,(\alpha=1,2,...,d)$ ($d$ is
the number of equations) satisfying
\begin{equation}
{\cal C}_\alpha: \,\,\,\displaystyle{\frac{dx}{dt}} = \lambda_{\alpha} \,\,\,\,\,(\alpha=1,2,...,d)
\end{equation}
\noindent
is said to be {\it genuinely nonlinear} or {\it linearly degenerate}
if, respectively,
\begin{eqnarray}
\label{GNLfield}
\vec{\nabla}_{\bf u} \,\, \lambda_{\alpha} \,\cdot\, {\bf r}_{\alpha} &\ne& 0\, ,\\ 
\vec{\nabla}_{\bf u} \,\, \lambda_{\alpha} \,\cdot\, {\bf r}_{\alpha} &=& 0 
\end{eqnarray}
\noindent
where the operator $\vec{\nabla}_{\bf u}$ acts on the space of conserved variables.

In a convex system, all the characteristic fields are genuinely
non-linear or linearly degenerate. Non-convexity is associated to
those states for which the condition (\ref{GNLfield}) is not fulfiled.

For the system of equations governing relativistic (ideal) flows it
can be shown that the convexity is strongly dependent on the second
derivatives of pressure (or the first derivatives of the sound
speed). In the following, we analize when genuinely nonlinear fields
become linearly degenerate, by examining the products ${\cal P}_{\pm}
:= \vec{\nabla}_{\bf w} \,\, \lambda_{\pm}({\bf w}) \cdot {\bf
  r}_{\pm}({\bf w})$. After some algebra, we find
\begin{equation}
{\cal P}_{\pm} =\,\pm\,\,T(a,b,c_s) \,\,
\left(\displaystyle{\frac{\partial c_s}{\partial \rho}} +
\displaystyle{\frac{p}{\rho^2}}\,\,
\displaystyle{\frac{\partial c_s}{\partial\epsilon}}
\,\, + \,\, \displaystyle{\frac{c_s}{\rho}}\, (1- c_s^2)\,\,
\right)
\label{Conv}
\end{equation}
being
\begin{equation}
T(a,b,c_s)
\,\, = \,\,
(a\,\, c_s \,\,\pm\,\, \delta^{1/2})^{-2}\,\,
(1-a^2)^2 \,\, \delta^{-1/2} \,\, W^2
\label{Conv1}
\end{equation}
where $a$ and $b$ stand for, respectively, the spatial components of
the velocity field in the x-direction and the tangential one. The
quantity $\delta$ is defined by $\delta = W^2 (1 - a^2- b^2c_s^2)$.

>From the above relations (\ref{Conv},\ref{Conv1}) it turns out that
the loss of convexity is closely related with the properties of the
equation of state (second term in Eq.~\ref{Conv}). The first and
second thermodinamical derivatives of pressure play a fundamental role
regarding with this issue (that was noticed by \cite{MP89}, for
equations of state having phase transitions). Furthermore, we realize that

\noindent
i) In the purely one-dimensional case ($b=0$), non-convexity only
appears in the ultrarelativistic regime ($a\rightarrow 1$):
\begin{equation}
  T(a,0,c_s) = 
  (1\, \pm \,a\,c_s)^{-2}\, W^{-2} \,
  \,\,
  ;\,\,\,\,
  a\, \rightarrow\, 1\, \Longrightarrow \,T(a,0,c_s)\, \longrightarrow\,  0 \,,\,{\cal{O}}\, (W^{-2})
\label{Conv2}
\end{equation}
ii) Likewise, if $a=0$, non-convexity arises in the
ultrarelativistic regime ($b\rightarrow 1$):
\begin{equation}
T(0,b,c_s) = 
\Delta^{-3/2} \,\, W^2
\,\,
;\,\,\,\,
b\, \rightarrow\, 1\, \Longrightarrow \,T(0,b,c_s)\, \longrightarrow\,  0 \,,\,{\cal{O}}\, (W^{-1})
\label{Conv3}
\end{equation}

\cite{BW88} noted that the equations of classical MHD are non-convex
at the degenerate states (magnetosonic waves change from genuinely
non-linear to linearly degenerate) . We have faced on the problem of
non-convex character of RMHD, and preliminar results allows one to
conclude that the degenerate states are, as in the classical MHD,
non-convex, being the magnetosonic fields the ones changing their 
character. We refer the reader to \cite{Anton08} (appendix G), 
where an analysis of the characteristic fields of RMHD in terms of Anile's 
covariant variables is presented. Much more theoretical work is necessary
in order to asses all the richness of other possible non-convex states in
RMHD.  The previous analysis in special relativistic hydrodynamics serves 
us as a road-map to the full characterization of RMHD pathological behaviours
(we remind the reader that the non-convex character of both the
classical and relativistic MHD equations is source of several
pathologies, as the development of the so-called compound waves).

%\acknowledgments 
{\bf Acknowledgments.\,\,\,}
Work supported by the grants AYA2007-67626-C03-01 and
CSD2007-00050 from the Spanish MICINN and PROMETEO/2009/103 of the
Generalitat Valenciana.

\bibliography{ms}

\begin{thebibliography}{}
\expandafter\ifx\csname natexlab\endcsname\relax\def\natexlab#1{#1}\fi
\expandafter\ifx\csname url\endcsname\relax
  \def\url#1{\texttt{#1}}\fi
\expandafter\ifx\csname urlprefix\endcsname\relax\def\urlprefix{URL }\fi
\providecommand{\eprint}[2][]{\url{#2}}

\bibitem[{Anile(1989)}]{anile}
Anile, A.~M. 1989, Relativistic fluids and magneto-fluids (Cambridge, England:
  Cambridge University Press)

\bibitem[{{Ant{\'o}n}(2008)}]{Anton08}
{Ant{\'o}n}, L. 2008, Ph.D. thesis, Universitat de Val{\`e}ncia

\bibitem[{{Ant{\'o}n} et~al.(2010){Ant{\'o}n}, {Miralles}, {Mart{\'{\i}}},
  {Ib{\'a}{\~n}ez}, {Aloy}, \& {Mimica}}]{Antonetal10}
{Ant{\'o}n}, L., {Miralles}, J.~A., {Mart{\'{\i}}}, J.~M., {Ib{\'a}{\~n}ez},
  J.~M., {Aloy}, M.~A., \& {Mimica}, P. 2010, \apjs, 188, 1

\bibitem[{Balsara(2001)}]{Balsara01}
Balsara, D. 2001, \apjs, 132, 83

\bibitem[{{Brio} \& {Wu}(1988)}]{BW88}
{Brio}, M., \& {Wu}, C.~C. 1988, J.~Comput.~Phys., 75, 400

\bibitem[{{Giacomazzo} \& {Rezzolla}(2006)}]{GR06}
{Giacomazzo}, B., \& {Rezzolla}, L. 2006, J.~Fluid~Mech., 562, 223

\bibitem[{{Koldoba} et~al.(2002){Koldoba}, {Kuznetsov}, \&
  {Ustyugova}}]{koldoba}
{Koldoba}, A.~V., {Kuznetsov}, O.~A., \& {Ustyugova}, G.~V. 2002, MNRAS, 333,
  932

\bibitem[{{Komissarov}(1999)}]{komissarov99}
{Komissarov}, S.~S. 1999, MNRAS, 303, 343

\bibitem[{{Mart{\'{\i}}} \& {M{\" u}ller}(2003)}]{martilr:03}
{Mart{\'{\i}}}, J.~M., \& {M{\" u}ller}, E. 2003, Living Rev.~Relativity, 6, 7

\bibitem[{{Menikoff} \& {Plohr}(1989)}]{MP89}
{Menikoff}, R., \& {Plohr}, B. 1989, Rev.~Mod.~Phys., 61, 75

\bibitem[{{Romero} et~al.(2005){Romero}, {Mart{\'{\i}}}, {Pons},
  {Ib{\'a}{\~n}ez}, \& {Miralles}}]{romero}
{Romero}, R., {Mart{\'{\i}}}, J.~M., {Pons}, J.~A., {Ib{\'a}{\~n}ez}, J.~M., \&
  {Miralles}, J.~A. 2005, J.~Fluid Mech., 544, 323

\end{thebibliography}
	
\end{document}